\documentclass[12pt]{article}
\usepackage{epsfig}
\makeatletter
\def\alt{\mathrel{\mathpalette\gl@align<}}
\def\agt{\mathrel{\mathpalette\gl@align>}}
\def\gl@align#1#2{\lower.6ex\vbox{\baselineskip\z@skip\lineskip\z@
\ialign{$\m@th#1\hfil##\hfil$\crcr#2\crcr\sim\crcr}}}
\makeatother

\begin{document}
\begin{flushright}
MIFP-07-15 \\
ACT-03-07\\
May, 2007
\end{flushright}
\vspace*{1.5cm}
\begin{center}
{
\Large\bf
No-Scale Solution to Little Hierarchy
}

\vspace{2cm}

{\large Bhaskar Dutta$^{1}$, Yukihiro Mimura$^{1}$ and Dimitri V.
Nanopoulos$^{1,2,3}$} \vspace{.5cm}

{\it $^{1}$Department of Physics, Texas A\&M University, College
Station,
TX 77843-4242, USA\\
\medskip
$^{2}$Astroparticle Physics Group, Houston Advanced Research Center
(HARC), \\ Mitchell Campus, Woodlands, TX 77381, USA\\
\medskip
$^{3}$Academy of Athens, Division of Natural Sciences, \\
28 Panepistimiou Avenue, Athens 10679, Greece } \vspace{.5cm}

\vspace{1.5cm}
{\bf Abstract}

\end{center}

We show that the little hierarchy problem can be solved in the
no-scale supergravity framework. In this model the supersymmetry
breaking scale is generated when the electroweak symmetry breaking
condition is satisfied and therefore, unlike usual supersymmetric
models, the correlation between the  electroweak symmetry breaking
scale
 and the average stop mass scale can be justified. This correlation
 solves the little hierarchy puzzle.
Using minimal supergravity boundary conditions,
 we find that the parameter space predicted by no-scale supergravity
 is allowed by all possible
 experimental constraints.
The predicted values of supersymmetric particle  masses are low
  enough to be very easily accessible at the LHC. This parameter space  will also be
  probed in the upcoming results from the dark matter direct detection
  experiments.

\thispagestyle{empty}

\bigskip
\newpage

\addtocounter{page}{-1}

\section{Introduction}
\baselineskip 18pt

Supersymmetry (SUSY) is one of the key ingredients to consider
physics beyond the Standard Model (SM). The large scale hierarchy
between the Planck scale and the weak scale is stabilized once the
hierarchy is generated.
In the minimal SUSY extension of the standard model (MSSM), the
electroweak symmetry breaking condition is satisfied
 by the renormalization
group running of the SUSY breaking mass for Higgs fields, and
therefore, a large hierarchy can be generated radiatively
\cite{Inoue:1982pi}: $m_W/ M_P \sim \exp(-4\pi^2)$.
In this picture we come across three different scales.  The scale
$Q_0$, 
where one of the eigenvalues of Higgs mass squared  becomes
negative, is much smaller than the Planck scale.
However, to generate the electroweak symmetry breaking vacua
radiatively 
 a typical SUSY breaking scale $Q_S$,
 where loop correction from the Higgs potential vanishes,
 is needed to be smaller
than the scale $Q_0$.
In addition, there is another scale,  $Q_{\rm st}$, where the
electroweak potential is destabilized in the $D$-flat direction. The
SUSY breaking scale $Q_S$ needs to be within the window between
$Q_{\rm st}$ and $Q_0$, i.e., $Q_{\rm st}<Q_S<Q_0$.
In the SUSY breaking models, the  scale $Q_S$ is just an input to
obtain a phenomenological model as an anthropic selection.

 The recent SUSY  particle search limits  seem to
demand an unnatural constraint on the radiative electroweak symmetry
breaking vacua to generate the correct $Z$ boson mass. The search
attempts have already exceeded the $Z$ boson mass scale. This means
that $Q_S$ is pushed up, and  a little hierarchy between the $Z$
boson mass and SUSY breaking masses gets created. Naively, if an
unnatural fine-tuning is not allowed, the electroweak symmetry
breaking condition leads to the fact that  $Q_S$, typically the
average stop mass,
 is not very large compared to  the $Z$ boson mass.
Surely, if we allow  fine-tuning, there is no problem. The
fine-tuning is encoded in the fact that  the two unrelated scales
$Q_S$ and $Q_0$ are close.
The scale $Q_0$ is obtained to be hierarchically small from the
Planck scale, and the hierarchy is determined by dimensionless
parameters. While the SUSY breaking scale $Q_S$ is a dimensionful
parameter of the model.
Why are two such unrelated scales destined to be close? Does there
exist any relation between $Q_0$ and $Q_S$? These are fundamental
questions and 
  require urgent
attention since the recent experimental constraints have caused a
little hierarchy between the $Z$ boson and SUSY breaking masses.

It is well known that SUSY is  an attractive candidate of physics
beyond the SM since it can solve the  unnatural tuning for the
quadratic divergence of Higgs mass. It also provides a dark matter
candidate, the neutralino, to explain the 23\% of the content of the
Universe~\cite{dark}. It is therefore important to understand
whether there exists a physics reason behind the selection of the
electroweak symmetry breaking vacua with little hierarchy.
%
One of the recent attempts is to reduce the fine-tuning in the
symmetry breaking condition by selecting
 a  SUSY breaking scenario
\cite{Choi:2005uz,Kitano:2005wc,Dermisek:2006ey}.
Another is to consider  statistically  probable vacua among the
electroweak symmetry breaking vacua
\cite{Giudice:2006sn,Dutta:2007az}. Such landscape idea can nicely
explain the little hierarchy. However, the selection of the
symmetric breaking vacua is still due to the  anthropic reason.

No-scale supergravity (SUGRA) model
\cite{Ellis:1983sf,Ellis:1983ei}, on the other hand, can explain not
only the selection of the electroweak symmetry breaking window, but
also the little hierarchy between the $Z$ boson mass and SUSY
breaking scale~\cite{Chikira:2000xi}.
In no-scale SUGRA, the gravitino mass is not determined due to the
flat potential and this continues until the gravitino mass or the
SUSY breaking feels the electroweak potential. The gravitino mass is
determined dynamically due to the radiative electroweak symmetry
breaking. In this sense, the radiative symmetry breaking vacua are
automatically selected. Therefore, the reason why  $Q_S$ is in the
symmetry breaking window is explained by its own mechanism.
Besides, the closeness of $Q_0$ and $Q_S$ is also realized by the
feature of  no-scale electroweak potential. So, the no-scale
structure is a golden solution of the little hierarchy problem.

No-scale SUGRA is well studied and has been well known for more than
twenty years \cite{Ellis:1983sf,Ellis:1983ei}. However, the no-scale
structure is often used only as a boundary condition at the
unification scale. In this paper, we discuss the no-scale structure
of the dynamical determination of the SUSY breaking scale as a
natural solution of the little hierarchy.
The electroweak symmetry breaking leads to two conditions
corresponding to the minimization by Higgs vacuum expectation values
(VEVs).
The dynamical determination of the SUSY breaking scale gives one
more relation between the $Z$ boson mass and the SUSY breaking
scale. The relation is written in terms of  the renormalization
group equations (RGEs) of the Higgs boson mass.
We describe the model constraints to generate radiative electroweak
symmetry breaking vacua, and find the prediction of  no-scale SUGRA.
Importantly, we find  that the SUSY breaking mass, typically the
stop and the gluino masses have upper bounds which are very easy to
reach at the upcoming collider experiments.
We also describe the phenomenological constraints and  show the
interesting prospect of discovering this model at the upcoming  dark
matter detection experiments.

This paper is organized as follows. In section 2, we discuss the
Higgs potential and see what kind of tuning is needed.
In section 3, we describe symmetry breaking vacuum
and no-scale SUGRA. In section 4, we discuss  no-scale supergravity
model and phenomenology.  Section 5 contains our conclusion.

\section{Higgs potential and Little Hierarchy}

The tree-level neutral Higgs potential is
\begin{equation}
V^{(0)} = m_1^2 v_d^2 + m_2^2 v_u^2 - (m_3^2 v_u v_d + c.c.)
+ \frac{g_Z^2}{8} (v_u^2-v_d^2)^2 ,
\end{equation}
where $v_u$ and $v_d$ are the VEVs of the neutral Higgs bosons,
$H_u^0$ and $H_d^0$.
The quartic coupling is obtained from $D$-term and
thus the coupling is related to the gauge couplings :
 $g_Z^2 = g^2+ g^{\prime2}$.
The quadratic terms are given by SUSY breaking Higgs masses,
$m_{H_d}^2$ and $m_{H_u}^2$, Higgsino mass $\mu$, and SUSY breaking
bilinear Higgs mass $B\mu$ : $m_1^2 = m_{H_d}^2 + \mu^2$, $m_2^2 =
m_{H_u}^2 + \mu^2$ and $m_3^2 = B\mu$.
The $Z$ boson mass is
$\frac{g_Z}{\sqrt{2}} v$, where $v = \sqrt{v_u^2+v_d^2}$. Minimizing
the tree-level Higgs potential (i.e., $\partial V/\partial v_u=0$,
$\partial V/\partial v_d=0$), we obtain
\begin{equation}
\frac{M_Z^2}2 = \frac{m_1^2 - m_2^2 \tan^2\beta}{\tan^2\beta -1},
\qquad
\sin2\beta = \frac{2 m_3^2}{m_1^2 + m_2^2},
\end{equation}
where $\tan \beta = v_u/v_d$.
The $Z$ boson mass can be also expressed as
\begin{equation}
\frac{M_Z^2}2 = -\mu^2 + \frac{m_{H_d}^2 - m_{H_u}^2 \tan^2\beta}{\tan^2\beta -1}
\equiv -\mu^2 + M_H^2.
\end{equation}
The SUSY breaking Higgs mass $M_H^2$ is approximately $-m_{H_u}^2$
for $\tan \beta \agt 10$. The electroweak symmetry can be broken by
RGE flow of the Higgs mass \cite{Inoue:1982pi}. Since the scale of
$M_H$ is naively governed by colored SUSY particles, it is not
comparable to the $Z$ boson mass using the current experimental
bounds on uncolored SUSY particles, if the universal boundary
condition is applied at the GUT or the Planck scale. Therefore,
fine-tuning is required between $\mu^2$ and $M_H^2$. So, naturalness
demands a model which generates smaller values of $\mu$ (corresponds
to smaller $M_H$) to reduce the fine-tuning \cite{Ellis:1986yg}.

Since the mass parameters run by RGEs, it is important to note the
scale where the fine-tuning is needed. Let us rewrite the expression
of the $Z$ boson mass to see what kind of tuning
is needed. 
The tree-level expression of $Z$ boson mass depends on scale $Q$,
and thus, let us define the $Q$ dependent $m_Z^2$,
\begin{equation}
m_Z^2 (Q) \equiv 2\frac{m_1^2(Q) - m_2^2(Q) \tan^2\beta(Q)}{\tan^2\beta(Q) -1}.
\end{equation}
Taking into account the 1-loop correction of the potential
\cite{Coleman:1973jx} in $\overline{\rm DR}^\prime$ scheme
\cite{Jack:1994rk},
\begin{equation}
V^{(1)} = \frac{1}{64 \pi^2}\sum_i (-1)^{2J_i}(2J_i +1) m_i^4
\left( \ln 
\frac{m_i^2}{Q^2}
-\frac32  \right),
\end{equation}
where $J_i$ is a spin of the particle $i$ with mass $m_i$,
we obtain
\begin{equation}
M_Z^2 = m_Z^2 (Q) + \frac1{v^2 \cos 2\beta}
\left(v_u \frac{\partial V^{(1)}}{\partial v_u}
-v_d \frac{\partial V^{(1)}}{\partial v_d}\right).
\label{Z-boson}
\end{equation}
This expression of $M_Z$ does not depend on $Q$ up to the wave
function renormalization for $v_u$ and $v_d$ at one-loop order.
Therefore the proper $Z$ boson mass is obtained approximately
at $Q = Q_S$
%
where
$\frac{\partial V^{(1)}}{\partial v_u}
=\cot \beta \frac{\partial V^{(1)}}{\partial v_d}$
is satisfied,
namely
\begin{equation}
e Q_S^2 = \prod_i (m_i^2)^{\frac{X_i}{X}},
\end{equation}
where
\begin{equation}
X_i = \left( \frac{\partial m_i^4}{\partial v_u} -
\cot \beta \frac{\partial m_i^4}{\partial v_d} \right) (-1)^{2 J_i} (2J_i+1),
\quad
X = \sum_i X_i.
\end{equation}
The scale $Q_S$ is naively the average of the stop masses.
Let us define the scale $Q_0$ where the function $m_Z^2(Q)$ is zero,
which is equivalent to the scale $m_1^2 m_2^2 = m_3^4$.
Then the $Z$ boson mass is expressed as
\begin{equation}
M_Z^2  \simeq \ln \frac{Q_S}{Q_0} \left.\frac{d m_Z^2}{d\ln Q}\right|_{Q= Q_0},
\end{equation}
and
\begin{equation}
\left.\frac{d m_Z^2}{d\ln Q}\right|_{Q= Q_0}
=
-\frac{2}{\cos^22\beta}
\left(
\frac{d m_2^2}{d\ln Q} \sin^2 \beta + \frac{d m_1^2}{d\ln Q} \cos^2 \beta
-                                     \frac{d m_3^2}{d\ln Q} \sin 2\beta
\right).
\end{equation}
For large $\tan \beta \agt 10$,
\begin{equation}
M_Z^2 \simeq \ln \left(\frac{Q_0}{Q_S}\right)^{\!2} \frac{d m_2^2}{d\ln Q}.
\label{Z-boson-2}
\end{equation}
{}From this expression, one can find that the $Z$ boson mass is
proportional to the stop mass up to a loop factor, and $Q_0$ and
$Q_S$ need to be close as needed by the little hierarchy between the
stop mass and $Z$ boson mass.
%
%
%
It is important to note that the smallness of the $\mu$ parameter is
not important in this expression since $\mu^2$ and $- m_{H_u}^2$ are
canceled in RGE at $Q=Q_0$\footnote{ Once $Q_S$ is fixed, the
smallness of $\mu$ (naturalness) is important for less fine-tuning
in the $Z$ boson mass. However, there is no reason that $Q_S$ is
fixed in general SUSY breaking model. If $Q_S$ is free, the tuning
parameter is $\ln Q_0/Q_S$, and then the smallness of $\mu$ is not
important for the tuning. }.
Therefore, the little hierarchy is characterized only by the
spectrum of stop masses in RGE of Higgs mass
and the closeness of $Q_0$ and $Q_S$.
For example, in the focus point solution \cite{Feng:1999mn} of
minimal supergravity (mSUGRA), it may give rise to a solution of the
naturalness problem if $Q_S$ is fixed at TeV scale (just below the
focus point) since the $\mu$ parameter is small. However, the little
hierarchy problem is not solved since the $Z$ boson mass is
sensitive to $\ln Q_0/Q_S$ and the stop masses are heavy in this
solution.

The radiative symmetry breaking elegantly explains the smallness of
$Q_0$ and the focus point scale compared to the Planck scale.
However, the hierarchy is determined irrespective of the overall
scale parameter since RGEs are homogenous differential equations,
and there is no reason that $Q_0$ and the focus point scales are
close to $Q_S$ (which is proportional to the overall scale). The
little hierarchy problem that we are  concerned about  is why such
unrelated scales are so close.

We can show that the closeness of $Q_0$ and $Q_S$  is probable among
the electroweak symmetry breaking vacua in the landscape picture
\cite{Dutta:2007az}.
However, in such a picture, the vacua where the electroweak symmetry
is not broken (namely $Q_0 < Q_S$) are also enormously probable, and
the electroweak symmetry breaking vacuum has a   special existence
among the multiverse. Obtaining the electroweak symmetry breaking
vacua could be just for anthropic reason at this stage.

In this paper, we stress that electroweak symmetry breaking vacua
with a little hierarchy is naturally obtained in no-scale SUGRA.

\section{Symmetry Breaking Vacuum and No-scale Model}

In this section, we study  the origin of the electroweak symmetry
breaking vacuum \cite{Ellis:1983sf} and the natural occurrence of the
closeness of $Q_0$ and $Q_S$ in no-scale electroweak potential.

In 
supergravity \cite{sugra,sugra1}, the SUSY breaking scale is
obtained in the hidden sector physics, and thus the scales $Q_0$ and
$Q_S$ are intuitively different and there is no reason that $Q_S$ is
selected in the electroweak symmetry breaking region.
In no-scale supergravity, on the other hand, the SUSY breaking scale
is not determined since the potential for the moduli $T$ and their
$F$-terms are completely flat. The SUSY breaking scale, which is a
function of $T$, is determined by the radiative effect of the Higgs
potential.
%
%
%
%
Since the dynamical determination of the SUSY breaking scale is due
to the electroweak radiative effect, $Q_0$ and $Q_S$ can be related
in the no-scale SUGRA.

The $Q$-independent electroweak potential is given as
\begin{equation}
V(v_u, v_d) = V^{(0)}(v_u,v_d;Q) + \Delta V(v_u,v_d;Q),
\end{equation}
where $\Delta V$ is loop correction and the Higgs VEVs-independent pieces
need to be subtracted,
\begin{equation}
\Delta V = V^{(1)} (v_u, v_d ; Q) - V^{(1)}(0,0; Q).
\end{equation}
When $Q_S$ is larger than $Q_0$, the electroweak symmetry does not
break, and thus $v_u= v_d =0$ and $V=0$.
If $Q_S$ is smaller than $Q_0$,
the $Q$-independent potential can be negative due to the
tree-level potential term.
In other words, at the minimal point of the
$Q$-independent potential $V(v_u,v_d,Q_S(T))$
(i.e., $\partial V/\partial v_u=0$, $\partial
V/\partial v_d=0$ and $\partial V/\partial T=0$),
the electroweak symmetry is broken.
Therefore, if there is no other hidden sector term to determine the scale $Q_S$,
the breaking condition $Q_0 > Q_S$ is
automatically satisfied 
in this framework. Besides, as we will see later, $Q_S$ is just
below the scale $Q_0$, and thus the scale $Q_S$ can be larger than
the stability-violating scale $Q_{\rm st}$.

Now let us consider a more concrete situation. We assume that every
mass parameter in the supergravity model is proportional to one mass
parameter (typically the gravitino mass). For example, in mSUGRA,
 the mass parameters  are $(m_0, m_{1/2}, A_0, \mu_0,
B_0)$, which are SUSY breaking scalar mass, gaugino mass, trilinear
scalar coupling, Higgsino mass and SUSY breaking bilinear Higgs mass
parameter, respectively.
Since the electroweak potential does not depend on gravitino mass
explicitly, it is useful to use the gaugino mass as an overall
scale.
A given no-scale model gives dimensionless parameters
$(\hat m_0, \hat A_0, \hat \mu_0, \hat B_0)$ and $\hat m_{3/2}$
e.g., $\hat m_0 = m_0/m_{1/2}$, $\hat A_{0} = A_{0}/m_{1/2}$, and so on.
The overall scale $m_{1/2}$ is determined by the electroweak
potential. In figure 1, we show the numerically calculated potential
minimized by $v_u$ and $v_d$ as a function of $m_{1/2}$ when $\hat
m_0 = \hat A_0 = \hat B_0 = 0$. The $\hat\mu_0$ parameter is chosen
to obtain the proper $Z$ boson mass at the minimum. In this choice,
$\tan\beta \sim 9$ at the minimal value of the potential.

\begin{figure}[t]
 \center
 \includegraphics[viewport = 20 20 270 220,width=8cm]{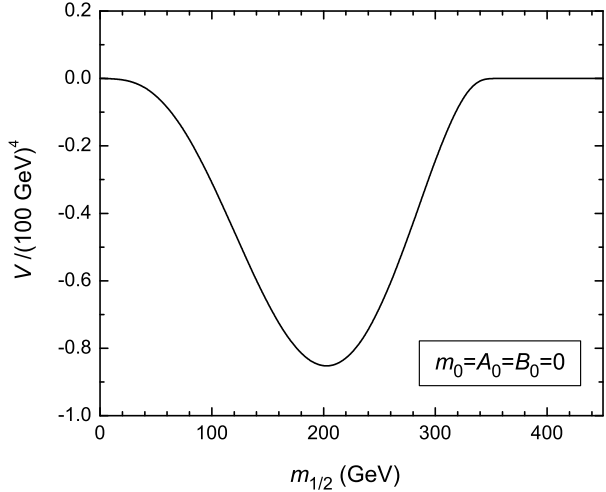}
 \caption{The no-scale electroweak potential}
\end{figure}

Now let us derive the fact that $Q_S$ and $Q_0$ are close at the
minimal point. The potential is obtained using the minimizing
conditions by $v_u$ and $v_d$ as
\begin{equation}
V = - \frac{1}{2 g_Z^2} M_Z^4 \cos^2 2\beta
+ \Delta V - \frac12 \left(v_u \frac{\partial \Delta V}{\partial v_u}
+ v_d \frac{\partial \Delta V}{\partial v_d} \right).
\end{equation}
Substituting Eq.(\ref{Z-boson}),
we obtain
\begin{equation}
V = - \frac1{2g_Z^2} m_Z^4 (Q) \cos^2 2\beta + \Delta V
- \frac{\sin2\beta}{2} \left(v_d \frac{\partial \Delta V}{\partial v_u}
+ v_u \frac{\partial \Delta V}{\partial v_d} \right) + \cdots .
\end{equation}
Since the potential is $Q$-independent, let us choose the scale $Q$
to make terms beyond the second term to be zero. Naively, it is the
scale where $\Delta V = 0$ when $\tan \beta$ is large. We call this
scale $Q_V$. The potential can be written as
\begin{equation}
V \simeq -\frac1{2g_Z^2} \left( \frac{d m_Z^2}{d \ln Q} \ln \frac{Q_V}{Q_0}
\right)^{\!2}\cos^2 2\beta .
\end{equation}
%
%
Since $\frac{d m_Z^2}{d \ln Q}$ is approximately proportional to the
overall scale which is related to $Q_V$, the potential is
\begin{equation}
V \propto - Q_V^4 \left(\ln \frac{Q_0}{Q_V}\right)^{\!2}.
\end{equation}
Minimizing the potential by $Q_V$, we obtain
$Q_V = Q_0/e^{1/2}$.
%
Thus the scale $Q_V$ is just below the symmetry breaking scale $Q_0$.
When we write $Q_S = k Q_V$, the $Z$ boson mass at the minimum is
obtained from Eq.(\ref{Z-boson-2})
\begin{equation}
M_Z^2 \simeq (1-\ln k^2) \left. \frac{d m_2^2}{d\ln Q} \right|_{Q=Q_0}.
\label{Z-boson-3}
\end{equation}
In the MSSM mass spectrum, the stop masses are important to
determine $Q_V$ and $Q_S$. Thus, these two scales are close and $k
\sim 1$. In the numerical calculation, $k$ depends on stop mixings
etc, but $\ln k^2$ is about $0.1 - 0.2$.
Note that the low energy particle spectrum ratio does not depend on
overall scale (we choose it as $m_{1/2}$),
when $(\hat m_0, \hat A_0, \hat \mu_0, \hat B_0)$ are fixed as
 boundary condition.
Therefore, $Q_V$ is naively proportional to $m_{1/2}$, and thus, the
minimization of the potential by an overall scale is rationalized.

The parameter $\hat \mu_0$ is consumed to fix $Z$ boson mass at the
minimum, and $\hat B_0$ is determined when $\tan \beta$ is fixed.
So, the model parameters in the minimal supergravity are $\hat m_0$,
$\hat A_0$, $\tan \beta$ and the signature of $\mu_0$.

Since the RGE of $m_2^2$ at $Q = Q_0$
is almost determined by stop mass parameters
with a loop factor,
\begin{equation}
\frac{d m_2^2 }{d\ln Q} \simeq \frac3{8\pi^2} (y_t^2 (m_{\tilde t_L}^2 + m_{\tilde t_R}^2)+A_t^2)
\end{equation}
the little hierarchy between the $Z$ boson and stop masses is
obtained by a minimization of the no-scale electroweak potential.
Numerically one finds that the gaugino mass at the  GUT scale is
about 200 GeV for small $\hat m_0$. This result does not depend on
$\tan\beta$ very much unless $\tan\beta$ is 
small.

\section{No-scale Model and Phenomenology }

In this section, we study the no-scale supergravity model
to realize the no-scale electroweak potential
in the previous section,
and find the phenomenological consequence of the model.
There are several ways to realize the no-scale structure
\cite{Ellis:1983sf,Ellis:1983ei}. Here, we consider the simplest
model to realize what we have described in the previous section.

In the simplest no-scale model, the K\"ahler potential is given
as~\cite{Ellis:1983ei}
\begin{equation}
{\cal G} = - 3 \ln (T+\bar T - \frac13 \phi_i \bar \phi_i) + \ln |W|^2,
\label{kahler0}
\end{equation}
where $\phi_i$'s are matter and Higgs fields. In this choice, $m_0$
and $A_0$ are zero as  boundary conditions. The $\mu$ term can be
proportional to gravitino mass when bilinear term $H_u H_d$ is in
the K\"ahler potential but not in the superpotential $W$.
More generally, one can write down the K\"ahler potential with
modular weights $\lambda_i$ and $\rho$ as
\cite{Ellis:1983ei,Ferrara:1994kg}
\begin{equation}
{\cal G} = -3 \ln (T+\bar T) + \frac{\phi_i \bar \phi_i}{(T+\bar T)^{\lambda_i}}
+ \frac{h}{2} \left( \frac{H_u H_d}{(T+\bar T)^\rho} + h.c. \right) + \ln |W|^2.
\end{equation}
Then, we obtain
\begin{eqnarray}
m^2_i &=& (1- \lambda_i) m_{3/2}^2,
\label{m0} \\
A_{ijk} &=& (3- \lambda_i - \lambda_j - \lambda_k) m_{3/2},
\label{A0} \\
B_0 &=& (2 - \lambda_{H_u} - \lambda_{H_d} + \rho) m_{3/2},
\end{eqnarray}
and $\mu$ term is proportional to $1-\rho$.
To make that the Higgsino mass $\mu$ is proportional to $m_{3/2}$,
we need $\lambda_{H_u}+\lambda_{H_d} = 2 \rho$.
The gravitino mass is $m_{3/2} = \frac{|W|}{M_P^2}\frac1{(T+\bar T)^{3/2}} $.
The modular weights $\lambda_i$ and $\rho$ are determined
in a concrete model \cite{Ferrara:1994kg,Camara:2003ku}.

The gauge kinetic function to determine the gaugino mass is
\begin{equation}
f_A = k_A T^{\ell_A}.
\end{equation}
In our assumption, every weak scale mass parameter is proportional
to one dimensionful mass. In order to achieve this, the gauge
kinetic function should depend only on the real part of $T$. Then
the modular weight $\ell_A$ needs to be 1 (or 0). Therefore, all
(kinetic normalized) gaugino masses are unified at the boundary,
while the gauge coupling constants can be different since $k_A$ can
be different. The gaugino mass is same as the gravitino mass at the
cutoff scale.

If there are fields which acquire heavy scale VEVs such as GUT Higgs
fields, these fields need to be inside the log as in
Eq.(\ref{kahler0}) so that the flat potential is not destabilized.

Even if the potential is flat at the tree-level, the quantum effects
may destroy the flatness \cite{Ferrara:1994kg}. The dangerous term
which destabilizes the electroweak scale is $\Lambda^2\, {\rm Str}
M^2/(32\pi^2)$, where $\Lambda$ is a cutoff scale. The supertrace is
proportional to $m_{3/2}^2$ and thus, it destroys the dynamical
determination of $m_{3/2}$ by electroweak potential. In a simplest
case, ${\rm Str} M^2$ is negative, and then the gravitino mass goes
to infinity. Therefore, ${\rm Str} M^2$ needs to be zero including
moduli and the hidden sector fields. Here after, we assume that the
supertrace is zero, which can be realized.

Though we have to forbid the $\Lambda^2 m_{3/2}^2$ term, there can
be a harmless correction such as $\alpha m_{3/2}^4$ term in the
potential. Such a term can arise due to Casimir effects which is
related to the SUSY breakings, or due to a correction in the
K\"ahler potential~\cite{Ferrara:1994kg}
\begin{equation}
- 3 \ln (T+\bar T) \rightarrow  -\ln ((T+\bar T)^3+c).
\end{equation}
When such a correction in the potential is taken into
account, the result in the previous section is modified.
The potential with the $\alpha m_{3/2}^4$ term is given, naively, as
\begin{equation}
V \propto - Q_V^4 \left(\ln \frac{Q_0}{Q_V} \right)^2 + \bar \alpha
Q_V^4,
\end{equation}
where $\bar\alpha$ is proportional to $\alpha$. Then, minimizing the
potential with respect to $Q_V$,
we obtain 
\begin{equation}
\ln \frac{Q_0}{Q_V} = \frac{1+\sqrt{1+16 \bar \alpha}}{4},
\end{equation}
and $\ln \frac{Q_0}{Q_V} > \frac14$ by using  $\frac{\partial^2
V}{\partial Q_V^2} >0$.
Therefore, we write
\begin{equation}
M_Z^2 \agt \frac12 (1-\ln k^2) \left. \frac{d m_2^2}{d\ln Q} \right|_{Q=Q_0},
\label{Z-boson-4}
\end{equation}
which provides an upper bound of the overall SUSY breaking scale
($m_{1/2}$) for given $\hat m_0, \hat A_0$, and $\tan \beta$.
The upper bound of the gaugino mass at the minimum is about $\sqrt
2$ times compared to the $\alpha = 0$ case Eq.(\ref{Z-boson-3}).
%
In figure 2, we show the numerical result of the minimization of
$m_{1/2}$ with the  experimental constraints.
We emphasize that the no-scale bound we have obtained does not
depend on the detail of the no-scale model constructed from string
theory. We obtain the no-scale bound, as long as there is no a
priori scale around the weak scale and the potential is flat.

\begin{figure}[t]
 \center
 \includegraphics[viewport = 0 90 535 655,width=8cm]{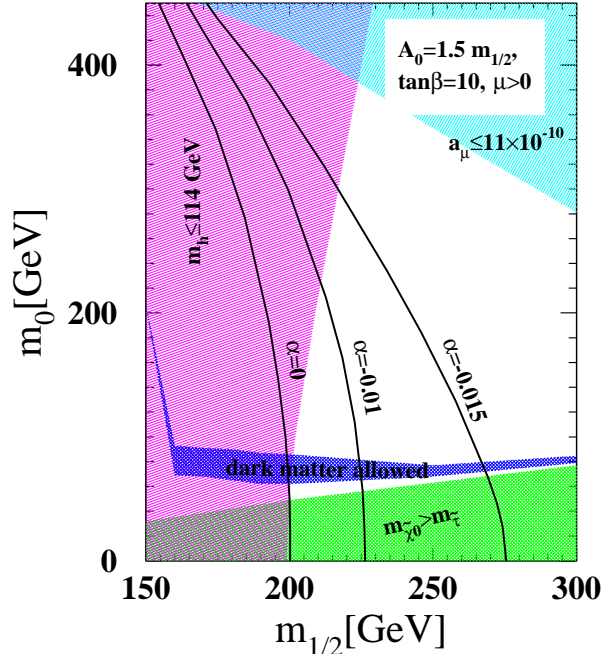}
 \caption{We show minimization contours of the potential for different values of $\alpha$ (defined
 in the text) in the mSUGRA parameter space. The blue narrow bands are allowed by dark matter
 constraints. The lightest Higgs mass $m_H\leq 114$ GeV is in the pink shaded region.
 $a_{\mu}\leq 11\times 10^{-10}$ in the light blue shaded region.}
\end{figure}

In drawing the figure, we assume universal scalar mass $m_0$ and
universal trilinear coupling $A_0$ at the GUT scale $\sim 2\times
10^{16}$ GeV. We use 2-loop RGE between the GUT and the weak scale
to determine the weak scale spectrum and 1-loop corrected potential
for minimization. The 1-loop potential has a slight $Q$-dependence
and it may change the result by a few percent. We choose the
evaluation scale $Q$ to be about 500 GeV so that the result is
insensitive to $Q$. If the SUSY breaking terms are universal, $m_0$
and $A_0$ are related, $\hat A_0 = 3 \hat m_0^2$, due to
Eqs.(\ref{m0},\ref{A0}), but we do not assume such relations in
drawing the figure because of the reason which we will describe
later.

As we have noted, the parameters are $\hat m_0$, $\hat A_0$, $\tan
\beta$ and also a signature for $\mu$. We choose $\mu>0$ due to
$b\to s \gamma$ constraint. We show the case for $\tan \beta = 10$
at the minimal point of the potential so that the region is allowed
by the Br[$b\to s \gamma$] which we take to be $2.2\times
10^{-4}<{\rm Br}[b\rightarrow s\gamma]<4.5\times
10^{-4}$~\cite{bsg}.
We choose $\hat A_0 = 1.5$ to satisfy the bound on the lightest
Higgs boson mass, as well as the Br$[b\to s \gamma]$ constraint.
Then, by changing $\hat m_0$, we obtain $m_{1/2}$ by numerical
minimization of the potential, and the solid lines are drawn.
The three lines corresponds to the value $\alpha = 0, -0.01,
-0.015$, where $\alpha$ is the coefficient of the correction of the
potential $\alpha m_{3/2}^4$. When $\alpha>0$, the $m_{1/2}$ value
at the minimal point of the potential  becomes smaller.
As one can understand from the discussion above, the minimal value
of $m_{1/2}$ is almost determined by the ratio $m_{1/2}/Q_V$. The
ratio is determined by the low energy spectrum, typically by stop
mass. Therefore, the solid lines are naively obtained by the
trajectories for constant average stop mass, and thus, they are
elliptic curves in the $m_0$-$m_{1/2}$ plain.
The solid lines are insensitive to the top quark mass, but depends
on strong gauge coupling $\alpha_3$. We use $\alpha_3
(M_Z)^{\overline{\rm MS}} = 0.117$, and $m_t = 172.7$ GeV.

In the figure, we also draw the experimental constraints for the
lightest Higgs boson mass~\cite{higgs}, muon anomalous magnetic
moment $a_\mu$~\cite{BNL}, dark matter \cite{WMAP}. We also show
the region where neutralino is the lightest SUSY particle.

It is interesting to note that the no-scale allowed region is within
the reach of the LHC, and in the mSUGRA model it is allowed by all
the experimental constraints. It is also important to note that the
dark matter $\tilde\chi^0_1$-$p$ cross-sections are (in $10^{-8}$
pb) 1.6-5, 1-2.7 and 0.3-1.8 for $\alpha=0$, $-0.01$ and $-0.015$
respectively. (The ranges in the cross-sections are obtained for the
experimental range of $\sigma_{\pi N}$, strange quark content of
proton and  strange quark mass \cite{Accomando:1999eg}). 
The recent upper limit on the
neutralino proton cross-section is $5.5\times 10^{-8}$ pb from the
XENON 10 experiment~\cite{xe10}. We see that the no-scale SUGRA
allowed region will be probed very soon in these direct detection
experiments.

The phenomenological constraints so far we have discussed are  for
mSUGRA models of soft SUSY breaking terms. Though we use the
universal boundary conditions for $m_0$ and $A_0$ for simplicity to
draw figure 2, the no-scale prediction does not depend on the detail
of the boundary conditions Eqs.(\ref{m0},\ref{A0}), as well as the
cutoff scale very much because the prediction is determined by the
low energy stop mass spectrum via
Eqs.(\ref{Z-boson-3},\ref{Z-boson-4}).
On the other hand, the experimental constraints depends on the
location of the cutoff scale as well as the universality conditions,
especially for the dark matter allowed region and the stau LSP
region.


The important prediction of the no-scale structure is
Eq.(\ref{Z-boson-3}) for true flat potential $\alpha=0$, and the
bound is obtained from Eq.(\ref{Z-boson-4}).
Eq.(\ref{Z-boson-4}) gives  upper bound to stop mass, and therefore,
generates upper bound to the gluino mass, as well. The gluino mass
is bounded as $m_{\tilde g} \alt 730$ GeV when the lightest Higgs
mass bound is taken into account. We show one example of the low
energy spectrum:
\begin{center}
\begin{tabular}{|c| c |c|c|c|c|c|c| }
\hline \hline parameters&\multicolumn{6}{c|}{masses in
GeV}\\\hline\hline $m_0$, $m_{1/2}$, $A_0$, $\tan\beta$&$m_{\tilde
g}$& $ m_{\tilde t_{1,2}}$&$m_{h,H}$
&$m_{\tilde\tau_{1,2}}$&$m_{\tilde\chi^0_{1,2,3,4}}$&$m_{\tilde\chi^{\pm}_{1,2}}$\\\hline
100, 300, 0, 10&730&480, 670&114, 450&151, 243&125, 233, 410,
430&233, 430\\\hline
\end{tabular}
\end{center}

Now we note the possible phenomenological solution in terms of the
modular weights of the no-scale model in the following. (1) Choose
the modular weight for slepton to be  1 in order not to overclose
the universe\footnote{There may be an exception when the Boltzmann
equation is modified \cite{dimitri}.},
and the cutoff scale is chosen to be around
a few times $10^{17}$ GeV to avoid stau LSP.
(2) Choose the modular weights for squarks and up-type Higgs field
to generate the  trilinear coupling with suitable values  to satisfy
the lightest Higgs boson mass bound and the $b\to s \gamma$
constraint especially for the case of true flat no-scale potential
$\alpha = 0$.
%

%

\section{Conclusion and Discussion}

Since the LEP and the Tevatron data do not show any direct evidence
for SUSY, the SUSY particle mass scale has become larger compared to
the $M_Z$ scale,
therefore a little hierarchy is created between this scale and the $Z$
boson mass scale.
In order to implement the little hierarchy, two apparently unrelated
scales $Q_0$ (where the electroweak symmetry breaks radiatively) and
$Q_S$ (the scale where the correct $Z$ boson mass gets produced)
need to be close satisfying $Q_S< Q_0$, and the closeness is
characterized by stop mass spectrum.
In this paper, we have investigated the no-scale SUSY breaking
models and found that the dynamical determination of the SUSY
breaking scale in these models provides a natural solution of the
little hierarchy. The two scales $Q_0$ and $Q_S$ get related in the
no-scale model since the electroweak symmetry breaking vacuum also
fixes  $Q_S$ as a minimal of the electroweak potential.
Since the potential, minimized by Higgs VEVs, is naively
proportional to $- m_Z^4$, a larger overall scale is favored and a
large $Q_S$ provides a small value for the potential.
However, when $Q_S$ becomes very close to $Q_0$ (which is
independent of the overall scale), the $Z$ boson mass becomes
smaller by definition and the potential becomes larger. As a result,
$Q_S$ is stabilized just below the scale $Q_0$.

We considered a no-scale potential where the potential is flat up to
the gravitino mass and assumed that all the  weak scale parameters
are proportional to a single scale, which is natural in no-scale
supergravity models. Then we found that the lighter stop and gluino
masses can be as large as  480  and 730 GeV respectively. These
masses can be easily accessed at the LHC. Further, the parameter
space is allowed by the Higgs mass bound and the Br[$b\rightarrow
s\gamma$] using the  mSUGRA boundary conditions. It is also
interesting to note that the dark matter detection cross section is
in the range from 0.3 to 5 $\times 10^{-8}$ pb. The future dark
matter detection experiments can easily probe these cross-sections.
The model also can be fit with proper modular weight factors for the
quark and lepton fields.

We now note what happens when we do not assume the single scale
proportionality factor for  the parameters. Suppose that both $\mu$
and $m_{1/2}$ are free and $\hat m_0$ and $\hat A_0$ are fixed. Then
$Q_0$  can be changed by varying $\mu$, while the scale $Q_H$, where
$M_H^2$ becomes zero, is independent of $\mu$ and $m_{1/2}$. As in
the case where $Q_0$ is fixed in the single scale proportionality,
$Q_S$ can be as large as $Q_H$ (but $Q_S < Q_H$). By definition,
$Q_S < Q_0 < Q_H$ is satisfied, and thus, all three scales are close
at the minimal value of the potential. The closeness of $Q_0$ and
$Q_H$ means that $\mu$ is small by definition and we find $\mu <
M_Z$
%
at the minimal point, which is already excluded by the chargino mass
bound.
When both $\mu$ and $B$ are free from the other SUSY breaking
parameters, one finds a non-stabilized direction to the $D$-flat
direction. Thus the $\mu$-$B$ ratio should be fixed in this case.
Therefore, the single scale proportionality is a rational assumption
in no-scale models.

In no-scale models, the potential for moduli $T$ is almost flat even
if we include the electroweak potential, and therefore, the mass of
the moduli is tiny,
%
%
i.e., $m_T \sim m_W^2/M_P \sim 10^{-5}$ eV.
This moduli mass $m_T \sim m_W^2/M_*$ (where $M_*$ is a fundamental
scale)  does not depend on the detail of the model when the no-scale
structure is broken by the radiative effect.
Such light moduli overclose universe if the  misaligned of the
moduli from its minimal value   is of the $O(M_P)$ after the
inflation \cite{Ellis:1987qs}.
In order to avoid this problem,  the misalignment from the minimal
value should be much less than the Planck scale \cite{Choi:1998dw}.
In other words, the moduli can be a part of dark matter. There are
other interesting cosmological implications of  no-scale model
\cite{Endo:2006xg}, which are out of scope of this paper.

Another point about the no-scale model is that the electroweak
potential is $-O(m_W^4)$ at the minimum. Therefore, we need to add
possible contribution to cancel the vacuum energy. Such contribution
can come from other stabilized moduli or hidden sector fields from
$F$ or $D$ term. However, such fields may destroy the no-scale
structure. In general, such contribution generates $m_X^{4-\gamma}
m_{3/2}^\gamma$ term in the potential. For example, we obtain
$\gamma = 4/3$ when we use a hidden field with modular weight 1.
When $m_X$ is around the weak scale, it can avoid the destabilization
of the no-scale electroweak potential and the vacuum energy can be
canceled. However, by $m_{3/2}$ minimization, such positive terms
require the overall scale to be smaller  which is disfavored by
experimental results. In order to make the model viable, we need to
make $\gamma =0$, which is possible from a $D$-term contribution,
 so that the correction to tune vacuum energy should not
depend at all on no-scale moduli $T$.

\section{Acknowledgments}

This work was supported in part by the DOE grant DE-FG02-95ER40917.



\begin{thebibliography}{99}
%
%
%
%
%
%
%
%



\bibitem{Inoue:1982pi}
K.~Inoue, A.~Kakuto, H.~Komatsu and S.~Takeshita,
Prog.\ Theor.\ Phys.\  {\bf 68}, 927 (1982);
%
Prog.\ Theor.\ Phys.\  {\bf 71}, 413 (1984);
%
L.~E.~Ibanez and G.~G.~Ross,
  Phys.\ Lett.\ B {\bf 110}, 215 (1982);
%
L.~E.~Ibanez,
Phys.\ Lett.\ B {\bf 118}, 73 (1982);
%
Nucl.\ Phys.\ B {\bf 218}, 514 (1983);
%
  J.~R.~Ellis, D.~V.~Nanopoulos and K.~Tamvakis,
  Phys.\ Lett.\  B {\bf 121}, 123 (1983);
%
  J.~R.~Ellis, J.~S.~Hagelin, D.~V.~Nanopoulos and K.~Tamvakis,
  Phys.\ Lett.\ B {\bf 125}, 275 (1983);
%
L.~Alvarez-Gaume, J.~Polchinski and M.~B.~Wise,
Nucl.\ Phys.\ B {\bf 221}, 495 (1983).





\bibitem{dark}H.~Goldberg,
  Phys.\ Rev.\ Lett.\  {\bf 50}, 1419 (1983);
  J.~R.~Ellis, J.~S.~Hagelin, D.~V.~Nanopoulos, K.~A.~Olive and M.~Srednicki,
  Nucl.\ Phys.\  B {\bf 238}, 453 (1984).


\bibitem{Choi:2005uz}
  K.~Choi, K.~S.~Jeong and K.~i.~Okumura,
  JHEP {\bf 0509}, 039 (2005)
  [hep-ph/0504037];
%
  K.~Choi, K.~S.~Jeong, T.~Kobayashi and K.~i.~Okumura,
  Phys.\ Lett.\ B {\bf 633}, 355 (2006)
  [hep-ph/0508029];
%
  hep-ph/0612258.



\bibitem{Kitano:2005wc}
  R.~Kitano and Y.~Nomura,
  Phys.\ Lett.\ B {\bf 631}, 58 (2005)
  [hep-ph/0509039];
%
  Phys.\ Lett.\ B {\bf 632}, 162 (2006)
  [hep-ph/0509221];
%
  Phys.\ Rev.\ D {\bf 73}, 095004 (2006)
  [hep-ph/0602096];
%
  hep-ph/0606134;
%
  Y.~Nomura and D.~Poland,
  Phys.\ Lett.\  B {\bf 648}, 213 (2007)
  [hep-ph/0611249].




\bibitem{Dermisek:2006ey}
  R.~Dermisek and H.~D.~Kim,
  Phys.\ Rev.\ Lett.\  {\bf 96}, 211803 (2006);
%
  R.~Dermisek, H.~D.~Kim and I.~W.~Kim,
  JHEP {\bf 0610}, 001 (2006).



\bibitem{Giudice:2006sn}
  G.~F.~Giudice and R.~Rattazzi,
  Nucl.\ Phys.\ B {\bf 757}, 19 (2006)
  [hep-ph/0606105].


\bibitem{Dutta:2007az}
  B.~Dutta and Y.~Mimura,
  Phys.\ Lett.\  B {\bf 648}, 357 (2007)
  [hep-ph/0702002].



\bibitem{Ellis:1983sf}
  E.~Cremmer, S.~Ferrara, C.~Kounnas and D.~V.~Nanopoulos,
  Phys.\ Lett.\  B {\bf 133}, 61 (1983);
%
J.~R.~Ellis, A.~B.~Lahanas, D.~V.~Nanopoulos and K.~Tamvakis,
  Phys.\ Lett.\  B {\bf 134}, 429 (1984).



\bibitem{Ellis:1983ei}
  J.~R.~Ellis, C.~Kounnas and D.~V.~Nanopoulos,
  Nucl.\ Phys.\  B {\bf 241}, 406 (1984);
%
  Nucl.\ Phys.\  B {\bf 247}, 373 (1984);
%
  Phys.\ Lett.\  B {\bf 143}, 410 (1984);
%
  A.~B.~Lahanas and D.~V.~Nanopoulos,
  Phys.\ Rept.\  {\bf 145}, 1 (1987).




\bibitem{Chikira:2000xi}
  Y.~Chikira and Y.~Mimura,
  hep-ph/0005231.





\bibitem{Ellis:1986yg}
  J.~R.~Ellis, K.~Enqvist, D.~V.~Nanopoulos and F.~Zwirner,
  Mod.\ Phys.\ Lett.\  A {\bf 1}, 57 (1986);
%
  R.~Barbieri and G.~F.~Giudice,
  Nucl.\ Phys.\  B {\bf 306}, 63 (1988);
%
  G.~W.~Anderson and D.~J.~Castano,
  Phys.\ Lett.\ B {\bf 347}, 300 (1995)
  [hep-ph/9409419];
%
  K.~L.~Chan, U.~Chattopadhyay and P.~Nath,
  Phys.\ Rev.\ D {\bf 58}, 096004 (1998)
  [hep-ph/9710473].


\bibitem{Coleman:1973jx}
  S.~R.~Coleman and E.~Weinberg,
  Phys.\ Rev.\ D {\bf 7}, 1888 (1973).

\bibitem{Jack:1994rk}
  I.~Jack, D.~R.~T.~Jones, S.~P.~Martin, M.~T.~Vaughn and Y.~Yamada,
  Phys.\ Rev.\  D {\bf 50}, 5481 (1994)
  [hep-ph/9407291].

%
\bibitem{Feng:1999mn}
  J.~L.~Feng, K.~T.~Matchev and T.~Moroi,
  Phys.\ Rev.\ Lett.\  {\bf 84}, 2322 (2000)
  [hep-ph/9908309];
%
  Phys.\ Rev.\  D {\bf 61}, 075005 (2000)
  [hep-ph/9909334].




\bibitem{sugra}
D.~Z.~Freedman, P.~van Nieuwenhuizen and S.~Ferrara,
  Phys.\ Rev.\ D {\bf 13}, 3214 (1976);
%
 S.~Deser and B.~Zumino,
  Phys.\ Lett.\ B {\bf 62}, 335 (1976);
%
A.~H.~Chamseddine, R.~Arnowitt and P.~Nath,
  Phys.\ Rev.\ Lett.\  {\bf 49}, 970 (1982).


\bibitem{sugra1}
R.~Barbieri, S.~Ferrara and C.~A.~Savoy,
  Phys.\ Lett.\ B {\bf 119}, 343 (1982);
%
L.~J.~Hall, J.~D.~Lykken and S.~Weinberg,
  Phys.\ Rev.\ D {\bf 27}, 2359 (1983);
%
P.~Nath, R.~Arnowitt and A.~H.~Chamseddine,
  Nucl.\ Phys.\ B {\bf 227}, 121 (1983);
%
 For a review, see 
  H.~P.~Nilles,
  Phys.\ Rept.\  {\bf 110}, 1 (1984).










\bibitem{Ferrara:1994kg}
  S.~Ferrara, C.~Kounnas and F.~Zwirner,
  Nucl.\ Phys.\  B {\bf 429}, 589 (1994)
  [hep-th/9405188].


%




\bibitem{Camara:2003ku}
  P.~G.~Camara, L.~E.~Ibanez and A.~M.~Uranga,
  Nucl.\ Phys.\ B {\bf 689}, 195 (2004)
  [hep-th/0311241];
%
  D.~Lust, S.~Reffert and S.~Stieberger,
  Nucl.\ Phys.\ B {\bf 706}, 3 (2005)
  [hep-th/0406092];
%
  Nucl.\ Phys.\ B {\bf 727}, 264 (2005)
  [hep-th/0410074];
%
  L.~E.~Ibanez,
  Phys.\ Rev.\ D {\bf 71}, 055005 (2005)
  [hep-ph/0408064];
%
  A.~Font and L.~E.~Ibanez,
  JHEP {\bf 0503}, 040 (2005)
  [hep-th/0412150].




\bibitem{bsg}
  M.~S.~Alam {\it et al.}  [CLEO Collaboration],
  Phys.\ Rev.\ Lett.\  {\bf 74}, 2885 (1995).




\bibitem{higgs}
  R.~Barate {\it et al.}  [LEP Working Group for Higgs boson searches],
  Phys.\ Lett.\ B {\bf 565}, 61 (2003);
%
  W.~M.~Yao {\it et al.}  [Particle Data Group],
  J.\ Phys.\ G {\bf 33}, 1 (2006).



\bibitem{BNL}
  G.~W.~Bennett {\it et al.}  [Muon $g-2$ Collaboration],
  Phys.\ Rev.\ Lett.\  {\bf 92}, 161802 (2004)
  [hep-ex/0401008];
S. Eidelman, Talk at ICHEP 2006, Moscow, Russia.




\bibitem{WMAP}
  D.~N.~Spergel {\it et al.}  [WMAP Collaboration], Astrophys.\ J.\ Suppl.\  {\bf 148}, 175 (2003)
  [astro-ph/0302209];
  astro-ph/0603449.
%


\bibitem{Accomando:1999eg}
  E.~Accomando, R.~Arnowitt, B.~Dutta and Y.~Santoso,
  Nucl.\ Phys.\  B {\bf 585}, 124 (2000)
  [hep-ph/0001019].




\bibitem{xe10} Talk by E. Aprile at APS 07, Jacksonville, USA.



\bibitem{dimitri}
  A.~B.~Lahanas, N.~E.~Mavromatos and D.~V.~Nanopoulos,
   hep-ph/0608153;
%
  Phys.\ Lett.\  B {\bf 649}, 83 (2007)
  [hep-ph/0612152].



\bibitem{Ellis:1987qs}
  J.~R.~Ellis, N.~C.~Tsamis and M.~B.~Voloshin,
  Phys.\ Lett.\  B {\bf 194}, 291 (1987).

\bibitem{Choi:1998dw}
  K.~Choi, E.~J.~Chun and H.~B.~Kim,
  Phys.\ Rev.\  D {\bf 58}, 046003 (1998)
  [hep-ph/9801280].



\bibitem{Endo:2006xg}
  M.~Endo, K.~Kadota, K.~A.~Olive, F.~Takahashi and T.~T.~Yanagida,
  JCAP {\bf 0702}, 018 (2007)
  [hep-ph/0612263].



\end{thebibliography}
\end{document}